# Randomness in appendage coordination facilitates strenuous ground self-righting

Qihan Xuan, *Chen Li

Department of Mechanical Engineering, Johns Hopkins University

*Corresponding author. Email: [redacted]

## Abstract

Randomness is common in biological and artificial systems, resulting either from stochasticity of the environment or noise in organisms or devices themselves. In locomotor control, randomness is typically considered a nuisance. For example, during dynamic walking, randomness in stochastic terrain leads to metastable dynamics, which must be mitigated to stabilize the system around limit cycles. Here, we studied whether randomness in motion is beneficial for strenuous locomotor tasks. Our study used robotic simulation modeling of strenuous, leg-assisted, winged ground self-righting observed in cockroaches, in which unusually large randomness in wing and leg motions is present. We developed a simplified simulation robot capable of generating similar self-righting behavior and varied the randomness level in wing-leg coordination. During each wing opening attempt, the more randomness added to the time delay between wing opening and leg swinging, the more likely it was for the naive robot (which did not know what coordination is best) to self-right within a finite time. Wing-leg coordination, measured by the phase between wing and leg oscillations, had a crucial impact on self-righting outcome. Without randomness, periodic wing and leg oscillations often limited the system to visit a few bad phases, leading to failure to escape from the metastable state. With randomness, the system explored phases thoroughly and had a better chance of encountering good phases to self-right. Our study complements previous work by demonstrating that randomness helps destabilize locomotor systems from being trapped in undesired metastable states, a situation common in strenuous locomotion.

## 1. Introduction

Randomness is common in biological systems, resulting either from stochasticity of the environment (Bovet and Benhamou, 1988) or noise in organisms themselves (for example, in sensing, information processing, and movement) (Faisal et al., 2008; Van Beers et al., 2002). Randomness is also common in man-made dynamical systems, which can come from sensors (How and Tillerson, 2001), information transfer (Nilsson et al., 1998), motor control (Ho, 1997), mechanical properties (Marti, 2003), and the environment (Byl and Tedrake, 2009). Typically, randomness degrades system performance and needs to be mitigated. For example, stochasticity in the terrain (surface slope variation) can drift the limit cycles of a passive dynamic walker, breaking the dynamic stability found in idealized, flat terrain and leading to metastable (locally attractive) behaviors (Byl and Tedrake, 2009). Neuromuscular noise also decreases walking stability (Roos and Dingwell, 2010). Sensory noise can overwhelm weak signals (Faisal et al., 2008) and compromise motion planning (How and Tillerson, 2001; Osborne et al., 2005). Inherent random time delay in communication and computation degrades the performance of control systems (Nilsson et al., 1998). All these problems can pose challenges to locomotion.

Although typically considered as a nuisance, randomness can be useful for both biological and artificial locomotor systems. Over large spatiotemporal scales, many animals move in stochastic trajectories (e.g., Lévy flight (Bénichou et al., 2005; Reynolds and Rhodes, 2009), correlated random walk (Bergman et al., 2000)) which increases the efficiency of searching for resources and mates (Reynolds and Rhodes, 2009) and decreases the risk of encountering predators (Bergman et al., 2000) or conspecifics that compete for the same resources (Reynolds and Rhodes, 2009). Over smaller spatiotemporal scales, stochasticity in the velocity of prey animals increases the probability of avoiding ballistic interception by predators (Moore et al., 2017). In biological sensing, weak periodical signals can be amplified via stochastic resonance with noise (under a threshold) (McDonnell and Ward, 2011; Wiesenfeld and Moss, 1995). Inspired by these biological systems, randomness has been leveraged to improve the performance of artificial systems, such as random search for optimization (Sutantyo et al., 2010), weak signal detection using stochastic resonance (Kurita et al., 2013) or resonant trapping (Gammaitoni and Bulsara, 2002).

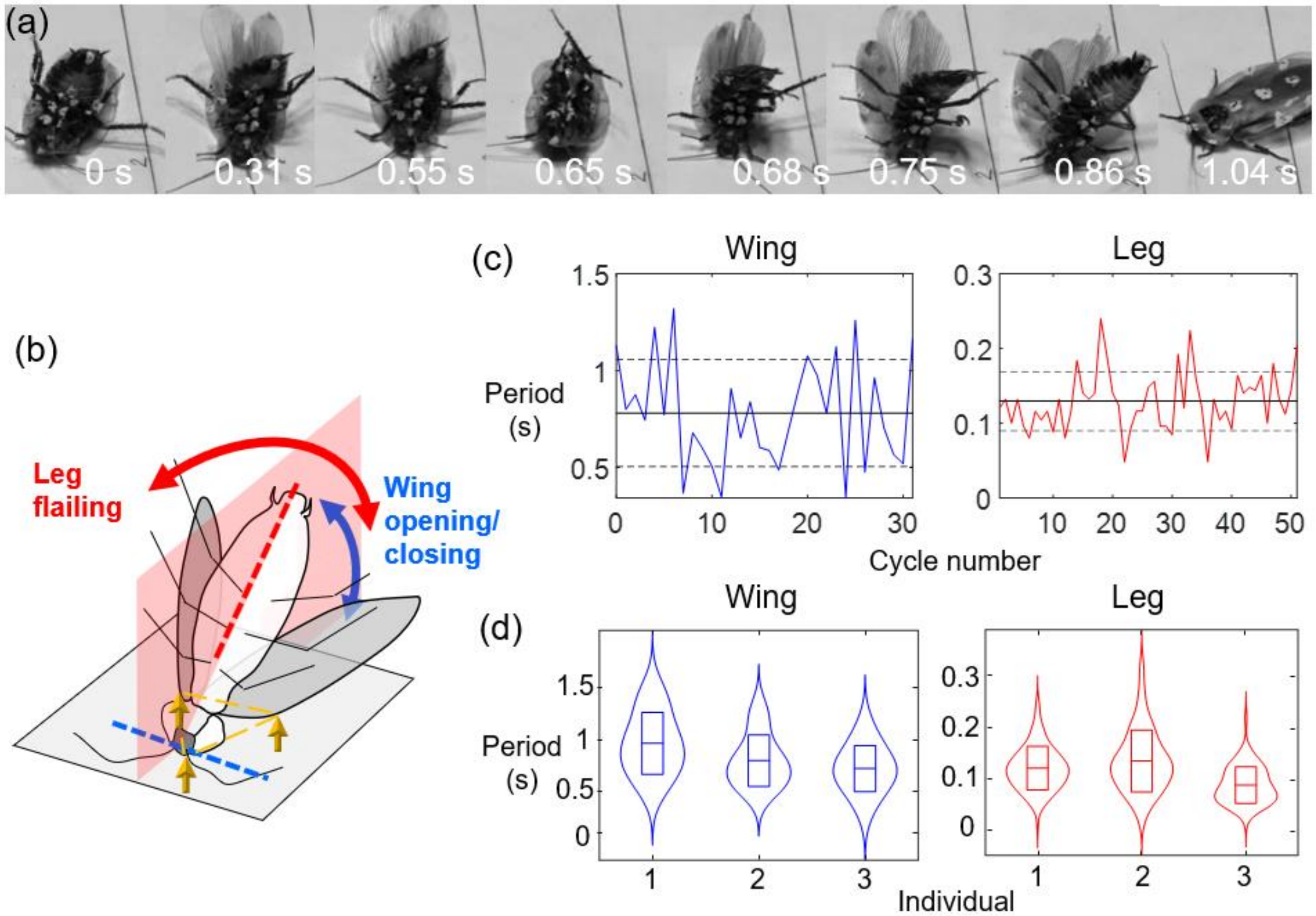

**Figure 1.** Discoid cockroach's strenuous, leg-assisted, winged ground self-righting. (a) Snapshots of a discoid cockroach self-righting. In this trial, the animal succeeds after two attempts of wing opening. (b) Schematic of animal in metastable state. Yellow arrows and triangle show head and two wings in contact with ground, forming a triangular base of support. Blue and red arrows show wing and leg oscillations. Blue and red dashed lines show approximate axes of rotation of body during wing oscillation and legs during flailing. Translucent red plane shows sagittal plane. (c) Left: periods of wing oscillation in a trial with 32 cycles. Right: periods of a hind leg's oscillation in a trial with 51 cycles. Solid and dashed lines show mean ± s.d. (d) Violin plots of wing and leg oscillation periods for three individuals ($n$ = 3 trials for each; 679 leg cycles and 59 wing cycles in total). Local width of graph shows the frequency of data along each value of y-axis. Inner rectangle shows mean ± s.d.

Inspired by these ideas, here we studied whether randomness in motion is beneficial for strenuous locomotor tasks. Our study was motivated by recent observation of the discoid cockroach (*Blaberus discoidalis*) self-righting from an upside-down orientation on a level, flat surface (figure 1(a), supplementary video S1) (Li et al., 2019). One strategy of the animal is to push both its wings together against the ground in an attempt to pitch over its head (figure 1(b), blue arrow) (Li et al., 2019). However, such a somersault has a large potential energy barrier, which the animal can rarely generate sufficient kinetic energy to overcome (Othayoth et al., 2017). Thus, the animal is often trapped in a metastable state (Hanggi, 1986), where its center of mass projection falls within a triangular base of support formed by the head and outer edges of two wings in contact with the ground (figure 1(b), yellow arrows and triangle) (Othayoth et al., 2017). Meanwhile, the animal often flails its legs laterally (figure 1(b), red arrow) and flexes and twists its abdomen (Li et al., 2019). These motions induce kinetic energy fluctuation to perturb the body to roll, which overcomes a smaller potential energy barrier (Othayoth et al., 2017). Thus, when the animal does eventually self-right, often after multiple wing opening attempts, it almost always rolls to one side (Li et al., 2019).

During each wing opening attempt, while mechanical energy is injected by wing and leg motions, it is dissipated via collision and friction against the ground and internal collision (wings and legs stop moving relative to the body). Thus, coordination between wing and leg oscillations may be critical for self-righting. Curiously, compared to cockroach walking (Watson and Ritzmann, 1997) and running (Full and Tu, 1990), in strenuous, leg-assisted, winged self-righting, both wing and leg oscillations are much less periodic, with large randomness present in their amplitudes, directions, speeds, and periods. For example, the periods of leg and wing oscillations are highly variable, with a coefficient of variation (standard deviation divided by mean) of $C_v = 25\%$ for leg and $C_v = 36\%$ for wing (figure 1(c, d)). In addition, the animal appears to randomly flex and twist its abdomen and scrape or hit its flailing legs against the ground by chance (Li et al., 2019). All these large random motions are absent during walking and running.

We hypothesize that the unusually large randomness in motions is beneficial for strenuous, leg-assisted, winged ground self-righting, by allowing random search in appendage configuration space to find an appropriate appendage coordination. We chose to focus on leg-assisted, winged self-righting among the diverse strategies observed in the cockroaches (Li et al., 2019), because it is a strenuous behavior where coordination between different appendages is likely to be critical. As described above, this self-righting behavior is complex, with the head, two wings, multiple legs, and abdomen all playing a role (Li et al., 2019). Studying randomness in the coordination of all these body parts together poses a significant challenge. As a first step, we focused on the randomness in the coordination between wings and legs.

To test our hypothesis, we took an approach of robotic modeling and created a simplified computational model—a cockroach-inspired simulation robot—to perform systematic in silico experiments (figure 2(b)). We chose to use a simulation robot here because it allowed precise control of randomness and large-scale, systematic parameter variation not practical in physical experiments. Our simulation robot followed the design and control of a recent physical robot that we developed to understand the role of kinetic energy fluctuation in leg-assisted, winged self-righting (Othayoth et al., 2017). The robot has two wings, a pendulum "leg", and a body with a head that protrudes beyond the anterior end of the wings (mimicking the cockroach's head). The two wings open and close symmetrically, and the single leg swings side to side to generate lateral perturbation. Although these motions are much simplified relative to the animal's, they can generate body motions representative of the animal's strenuous self-righting behavior that we are interested in (Othayoth et al., 2017), while also providing the simplest model system, with only two degrees of freedom in actuation.

As a first step to understand the role of randomness in wing-leg coordination, we added Gaussian noise of variable levels to wing oscillation period (figure 2(c)). During each wing oscillation cycle, between wing opening and closing, the wings are held open or closed for some time (supplementary video S1). We chose to only add noise to the time that wings are held closed, so that other parameters (wing opening/closing speed, wing opening amplitudes, wing opening time) were kept constant. In addition, we varied wing opening and leg flailing amplitudes (figure 6) to study the effect of randomness over a wide range of parameter space. Then, we studied how randomness in the phase between wing and leg oscillations affected self-righting outcome of a single wing opening attempt, and we used this single attempt phase map (figure 7) to predict the outcome of multiple attempts (figure 8). These analyses helped reveal how randomness in phase was beneficial for self-righting.

We emphasize that we deliberately designed and controlled our robot to serve as a physical model to generate strenuous self-righting similar to the animal's, so that we could study appendage coordination. Our goal is not to simply achieve successful self-righting, which can be done in many other, and often simpler, ways in a robot (for a review, see (Li et al., 2017)). For example, a previous cockroach-inspired robot with wings and no legs (Li et al., 2017) was capable of self-righting by a somersault using wings only. It could also open the left and right wings asymmetrically to roll the body to self-right. In both cases, the wings opened sufficiently to generate sufficient rotational kinetic energy to overcome the potential energy barrier. By contrast, the discoid cockroach's wing opening alone was rarely sufficient (Othayoth et al., 2017) and must be supplemented by the perturbating motions of the legs and abdomen. In addition, this previous robot did not have a protruding head that adds to the potential energy barrier and makes self-righting strenuous. Thus, although this previous robot self-rights easily, it is not suitable for studying the behavior we are interested in here.

## 2. Methods

### 2.1. Design and actuation of simulation robot

Our simulation robot was created using Chrono, an open-source, high-fidelity, multi-body dynamics engine (Mazhar et al., 2013; Tasora et al., 2015). Besides the head, two wings, single pendulum leg (consisting of a lightweight rod and an added mass), the robot also had five cuboidal motors (figure 2(a); see mass distribution in Table 1). The wings and head were cut from a thin ellipsoidal shell. We carefully matched the simulation robot's geometry and mass distribution. We created CAD models of these parts in SolidWorks and assembled them using the relative position and orientation of each part measured from the physical robot. We then exported the assembled CAD model into Chrono to create the simulation robot. We rounded the edges of each part to make the contact forces change more smoothly in simulation and considering that we used the Hertzian contact model (see below) which is developed for rounded shapes.

**Table 1.** Mass distribution of the simulation robot.

| Component | Mass (g) |
|---|---|
| Head | 13.4 |
| Leg rod | 4.3 |
| Leg added mass | 51.5 |
| Leg motor | 28.6 |
| Two wings | 57.4 |
| Two wing pitch motors | 56.0 |
| Two wing roll motors | 48.8 |
| Total | 260 |

Each wing could both pitch and roll relative to the body, actuated by two motors with orthogonal axes of rotation (figure 2(a)). We defined wing pitch and roll (figure 2(b)) as the angles rotated by the pitch motors (3, 4) and roll motors (1, 2). Because the cockroach's two wings open and close together during self-righting, we controlled the robot's two wings to move symmetrically. We also constrained wing pitch and roll angles to always be the same to simplify experiments and analysis. Thus, wings motion was effectively one degree of freedom, which we described with wing angle $\theta_w$ (figure 2(b)). The robot's pendulum leg was actuated by a separate motor (figure 2(a), motor 5). We defined leg angle $\theta_l$ as the angle between the pendulum and body midline (figure 2(b)).

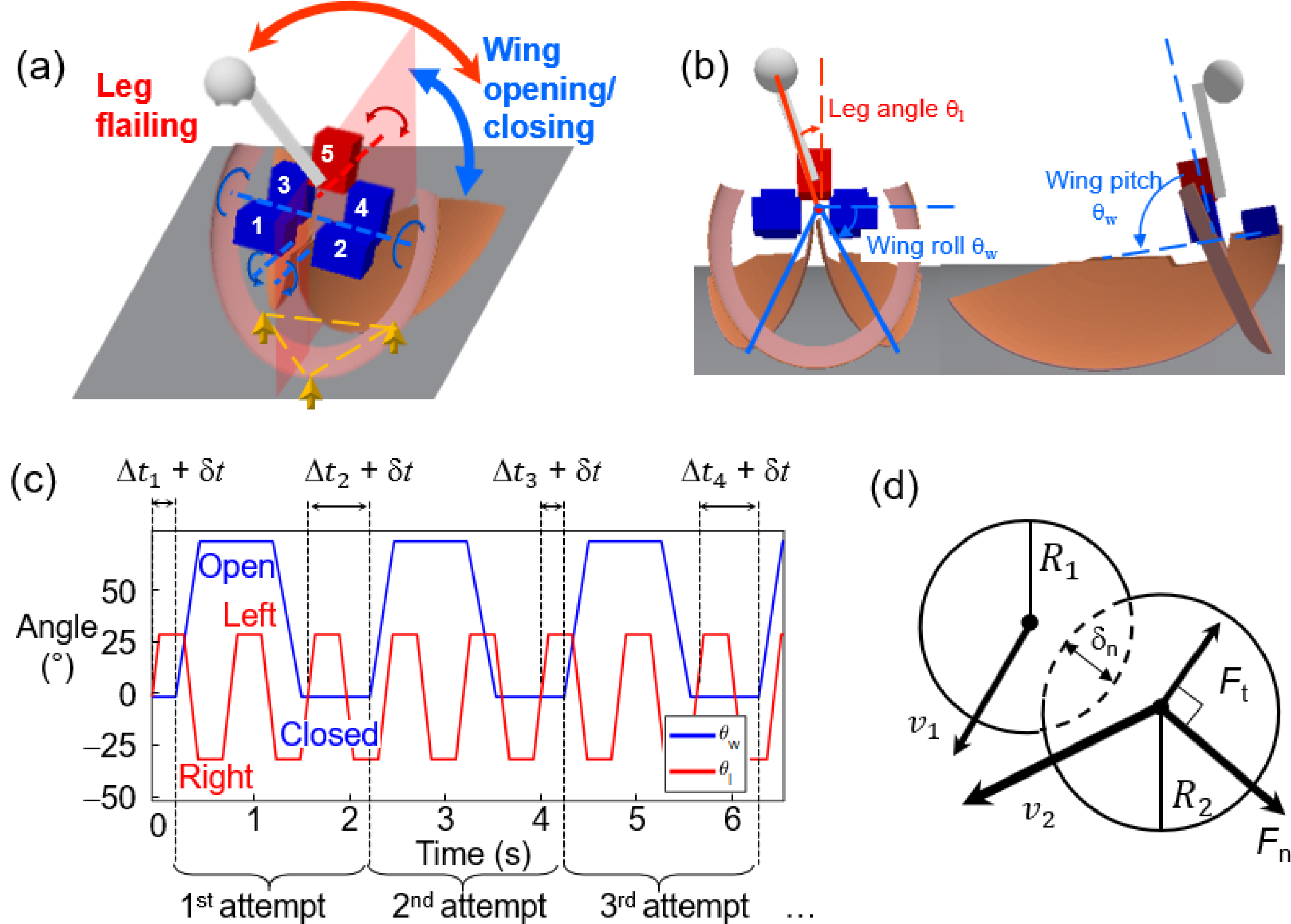

**Figure 2. Cockroach-inspired, leg-assisted, winged self-righting simulation robot.** (a) Simulation robot with a head, two wings, five motors, and a pendulum leg, in metastable state. Yellow arrows and triangle show head and two wings in contact with ground, forming a triangular base of support. Red and blue arrows show wing and leg oscillations. Translucent red plane shows sagittal plane. (b) Frontal and side views of simulation robot to define leg angle (red) as well as wing roll and wing pitch (blue). (c) Actuation profiles of wings (blue) and leg (red). $\Delta t_i$ is the time delay of the i$^{th}$ wing opening attempt, defined as the time interval between wing opening moment and the start of the preceding leg oscillation. A Gaussian noise $\delta t$ is added to $\Delta t$ in simulation experiments with randomness. (d) Hertzian contact model used in multi-body dynamics simulation. $\delta_n$ is virtual overlap (deformation) between two rigid bodies. $F_n$ and $F_t$ are normal and tangential contact forces. $R_1$ and $R_2$ are local radii of curvature at contact.

Our physical robot's wing and leg oscillations were controlled using simple actuation profiles, with actuation parameters deliberately chosen to generate strenuous leg-assisted, winged self-righting behavior similar to the animal's (Othayoth et al., 2017). Thus, we simply designed the simulation robot's wing and leg actuation profiles (figure 2(c)) to approximate those of the physical robot and used similar actuation parameters. For both the physical and simulation robots, wing and leg oscillation periods were $T_w$ = 2 s and $T_l$ = 0.8 s (except in phase-based prediction where we varied $T_w$ and $T_l$, see Section 3.6). $T_w$ was chosen to be greater than $T_l$ as observed in the animal. Although there may be an optimal frequency (close to the natural frequency in the body roll direction) for leg flailing to induce resonance, we did not study it here because we focused on the randomness in wing-leg coordination. For the simulation robot, wing opening and closing speeds were 300 °/s and 250 °/s, and the time for the leg to move from one side to the other was $t_l$ = 0.15 s. Thus, the angular speed of leg rotation was $2\theta_l/t_l$. Note that these were slightly different from the physical robot, whose values were 266 ± 19 °/s, 375 ± 14 °/s, and 0.143 ± 0.038 s. In particular, the slower wing closing speed in simulation was chosen to better match the physical robot. The physical robot's thin wings and head were deformable, and its 3-D printed plastic joints between the body and wings had a slight give under load, both of which quickly damped out body oscillation on the ground after the wings closed. The simulation robot's thin wings and head were rigid, and slower wing closing reduced body oscillation and simulated this effect. In physical/simulation robot experiments, we defined the time interval between two consecutive instants when the wings began to open as one wing opening attempt.

### 2.2. Contact mechanics model

To solve for dynamics, we used the discrete-element method via penalty (DEM-P) in Chrono, which models contact by a viscoelastic force model (Fleischmann, 2015; Tasora et al., 2015) (figure 2(d)):

$$\begin{cases} \boldsymbol{F}_n = f(\bar{R}, \delta_n)(k_n \boldsymbol{\delta}_n - \gamma_n \bar{m} \boldsymbol{v}_n) \\ \boldsymbol{F}_t = f(\bar{R}, \delta_n)(-k_t \boldsymbol{\delta}_t - \gamma_t \bar{m} \boldsymbol{v}_t) \end{cases} \quad (1)$$

where $\boldsymbol{\delta} = \boldsymbol{\delta}_n + \boldsymbol{\delta}_t$ is the displacement vector of the contact point between two bodies, which represents their overlap if no deformation occurs, and the subscripts "n" and "t" represent normal and tangent components, respectively. The scalars $\bar{m} = m_1 m_2/(m_1 + m_2)$ and $\bar{R} = R_1 R_2/(R_1 + R_2)$ are the effective mass and effective radius of curvature of the two interacting bodies. The vector $\boldsymbol{v} = \boldsymbol{v}_n + \boldsymbol{v}_t$ is the relative velocity between the two bodies. $k_n, k_t, \gamma_n, \gamma_t$ are the normal and tangential stiffness and damping coefficients, all of which depend on $\overline{m}$, $\overline{R}$, and material properties (Young's modulus $E$, Poisson's ration ν, and coefficient of restitution CoR) of the two bodies. Assuming Coulomb friction (with coefficient μ), sliding happens if $|\boldsymbol{F}_t| > \mu|\boldsymbol{F}_n|$ and stops otherwise. During sliding, kinetic friction is set to be $|\boldsymbol{F}_t| = \mu|\boldsymbol{F}_n|$. Here we chose Hertzian contact theory (Ding et al., 2012; Popov, 2010), i.e., in Eqn. 1:

$$f(\overline{R}, \delta_n) = \sqrt{\overline{R}\delta_n} \tag{2}$$

**2.3. Material property characterization**

To validate our simulation robot, we performed experiments to characterize material properties for the physical robot (Othayoth et al., 2017). This is important because the viscoelastic model (Eqn. 1) includes Young's modulus $E$, coefficient of friction μ, coefficient of restitution CoR, and Poisson's ratio ν as parameters. In physical robot experiments, before self-righting, only the wings and head of the robot (made of polystyrene) contact the ground (Othayoth et al., 2017). Thus, we measured or estimated the material property of polystyrene for model input.

To measure coefficient of friction, we set the physical robot upside down on a rigid plate with the sandpaper surface used in robot experiments (Othayoth et al., 2017). Then we slowly increased the slope angle α until the robot started to slip. The critical angle $\alpha_c$ was used to calculate the coefficient of friction via $\mu = \tan\alpha_c$. From this experiment (5 trials), $\mu = 1.00 \pm 0.04$ (mean ± s.d.). Thus, we used $\mu = 1$ in simulation.

To measure Young's modulus, we did extension experiments using an Instron universal testing machine (34TM-10, Norwood, MA, US). We tested three polystyrene beams with different thicknesses, with each beam tested five times. We found that $E = 0.78 \pm 0.04 \times 10^9$ Pa for polystyrene (mean ± s.d.). However, with such a high $E$, a simulation time step of $< 10^{-7}$ s was required for numerical convergence, leading to an impractical time to complete our simulation experiments (over one day to run 1 trial on a 3.4 GHz 16-core workstation). A common practical solution to this problem is to reduce Young's modulus in simulation so that numerical convergence can be achieved with a larger time step (Maladen et al., 2011; Pazouki et al., 2017; Tasora et al., 2015). We found that, with a sufficiently small simulation time step ($10^{-5}$ s), simulation results were not sensitive to $E$ within $10^5$ Pa to $10^7$ Pa (e.g., the measured self-righting time changed by less than 10%). Thus, we used a smaller $E = 1 \times 10^5$ Pa.

To obtain numerical convergence for this chosen $E$ while keeping simulation time practical, we varied time step from $10^{-3}$ s to $10^{-7}$ s. We found that using a larger time step of $10^{-4}$ s only resulted in a small error from using a smaller time step of $10^{-6}$ s (e.g., pitch, roll, and yaw angles of the robot changed less than 0.2° over 10 seconds of simulation). In addition, it reduced simulation time to yield a practical time to complete our simulation experiments (one day to run 1000 trials on a 3.4 GHz 16-core workstation). Thus, we used a time step of $10^{-4}$ s for all simulation experiments. The numerical convergence was not sensitive to other parameters ($\mu$, $\nu$, CoR), which only led to changes of smaller than an order of magnitude in the choice of time step.

In addition, we found that simulation results were insensitive to Poisson's ratio (e.g., pitch, roll, and yaw angles of the robot changed less than 0.5° over 10 seconds of simulation as $\nu$ varied from 0 to 0.5). Thus, we chose $\nu = 0.35$, close to that of polystyrene (Bangs Laboratories, 2015).

Because CoR was a function of the collision velocity, geometry, and material composition of both objects in contact (Ramírez et al., 1999), it was difficult to measure experimentally. We found that CoR had a small effect (e.g., self-righting time increased by 18.4% as CoR increased from 0 to 0.5). We chose CoR = 0.1 to achieve large dissipation in simulation to better match that the high damping of body oscillations of the physical robot on the ground after wings closed, as mentioned above.

**2.4. Validation of simulation against physical robot experiments**

We performed simulation experiments to verify that our simulation robot was reasonable in physics. In both physical and simulation robot experiments, we varied wing opening and leg oscillation amplitudes, $\theta_w$ and $\theta_l$, and measured self-righting time. The physical robot had a naturally occurring randomness in wing oscillation period of $C_v = 1.3\%$. For the simulation robot, we added a randomness of $C_v = 2.9\%$ to wing oscillation period. This was set to be larger than that in the physical robot to account for other randomness in the system, such as randomness in $\theta_w$ and $\theta_l$, leg actuation, and the environments. For both the simulation and physical robots, we performed five trials at each combination of $\theta_w$ and $\theta_l$. We recorded the first 10 seconds of each trial. The 10 seconds time limit was chosen to save simulation time, considering that in most trials, self-righting occurred within 10 seconds. If the robot could not self-right within 10 seconds, we defined it to have failed. For failed trials, we set self-righting time as 10 seconds. This was considering that, if we did not consider failed trials in averaging self-time, the few successful trials were not representative.

**2.5. Randomness in simulation robot motion**

To introduce randomness in wing oscillation period (and thus randomness in wing-leg coordination), we added Gaussian noise with variance $\sigma^2$ (using C++) to the time when wings are held closed. To isolate the effect of coordination, we chose to add randomness rather than vary $T_w$ and $T_l$ directly, because doing so would affect the mechanical energy injected by changing the speed and duration of motor actuation. To isolate the effect of randomness in actuation, we did not introduce noise in the mechanical system (e.g., morphology, physical property, the environment) that is inevitable in the physical robot and animal (see discussion in Section 4.2). For simulation experiments, we measured the level of randomness using coefficient of variation, $C_v$, defined as the ratio between standard deviation $\sigma$ and leg period $T_l$. We chose to normalize $\sigma$ by $T_l$ because phase $\varphi$, which we used to measure wing-leg coordination, was normalized by $T_l$. To study the effect of randomness, we varied $C_v$ (0%, 5%, 10%, 15%, 20%, and 25%) in simulation experiments. For each combination of $\theta_w$ (70°, 72°, 74°, 76°) and $\theta_l$ (20°, 30°, 40°, 50°) and a given $C_v$, we performed 40 simulation trials. This resulted in a total of 3840 trials.

**2.6. Phase between wing and leg oscillations**

Because we only added randomness to wing oscillation period while keeping everything else constant (Section 2.5), the only thing that changed was the phase between wing and leg oscillations at each wing opening attempt (figure 2(c)). For each wing opening attempt i, we defined the phase $\varphi$ between wing and leg oscillations as the ratio of time delay $\Delta t_i$ to leg period $T_l$. The initial phase was thus $\varphi_1 = \Delta t_1/ T_l$. To study the effect of phase for a single wing opening attempt (Section 3.5), we varied $\varphi$ from 0% to 100% with an increment of 5% for each combination of $\theta_w$ and $\theta_l$ tested, without adding randomness. To test the predictive power of phase-based prediction method (Section 3.6), we varied $\varphi$ from 0% to 100% with an increment 5% for each combination of wing period (2 s, 2.5 s, 3 s) and leg oscillation period (0.6 s, 0.8 s, 1 s, 1.2 s, 1.4 s) tested. We note that the animal's phase also varied from 0% to 100% (Appendix).

**2.7. Potential energy barrier calculation**

Self-righting requires overcoming a potential energy barrier (Domokos and Várkonyi, 2008). As a proxy to quantify the difficulty of self-righting over the range of $\theta_w$ and $\theta_l$ tested, we calculated how the minimal potential energy barrier to self-right changed with $\theta_w$ and $\theta_l$. When the wings were fully open, the center of mass (CoM) was at a local minimum on the potential energy landscape, which corresponded with the metastable state with triangular base of support (yellow arrows and triangle). If the robot pivots over an axis formed by the edge of head and a wing to self-right (figure 3), it overcomes a minimal potential energy barrier $\Delta E = mg\Delta h$. The simulation (and physical) robot did not always do so when self-righting, as its head or wing edge may lift off briefly during pivoting during dynamic rotation, and the actual barrier overcome may be slightly higher. However, the minimal barrier still provided a measure of how challenging it was to self-right.

To calculate the minimal barrier as a function of $\theta_w$ and $\theta_l$, we first calculated the gravitational potential energy landscape of the robot over body pitch and roll space (Othayoth et al., 2017), for the range of $\theta_w$ and $\theta_l$ tested above. We imported the robot CAD model in MATLAB and varied body pitch and roll from −180° to 180° for each combination of $\theta_w$ and $\theta_l$ to calculate its gravitational potential energy (we did not vary body yaw because it did not affect gravitational potential energy). Then, for each combination of $\theta_w$ and $\theta_l$, we searched for the minimal potential energy barrier using a Breadth-first search method. Note that the potential energy landscape here is different and an advancement over the simplistic landscape in previous studies (Li et al., 2017; Li et al., 2019), which only considered the body.

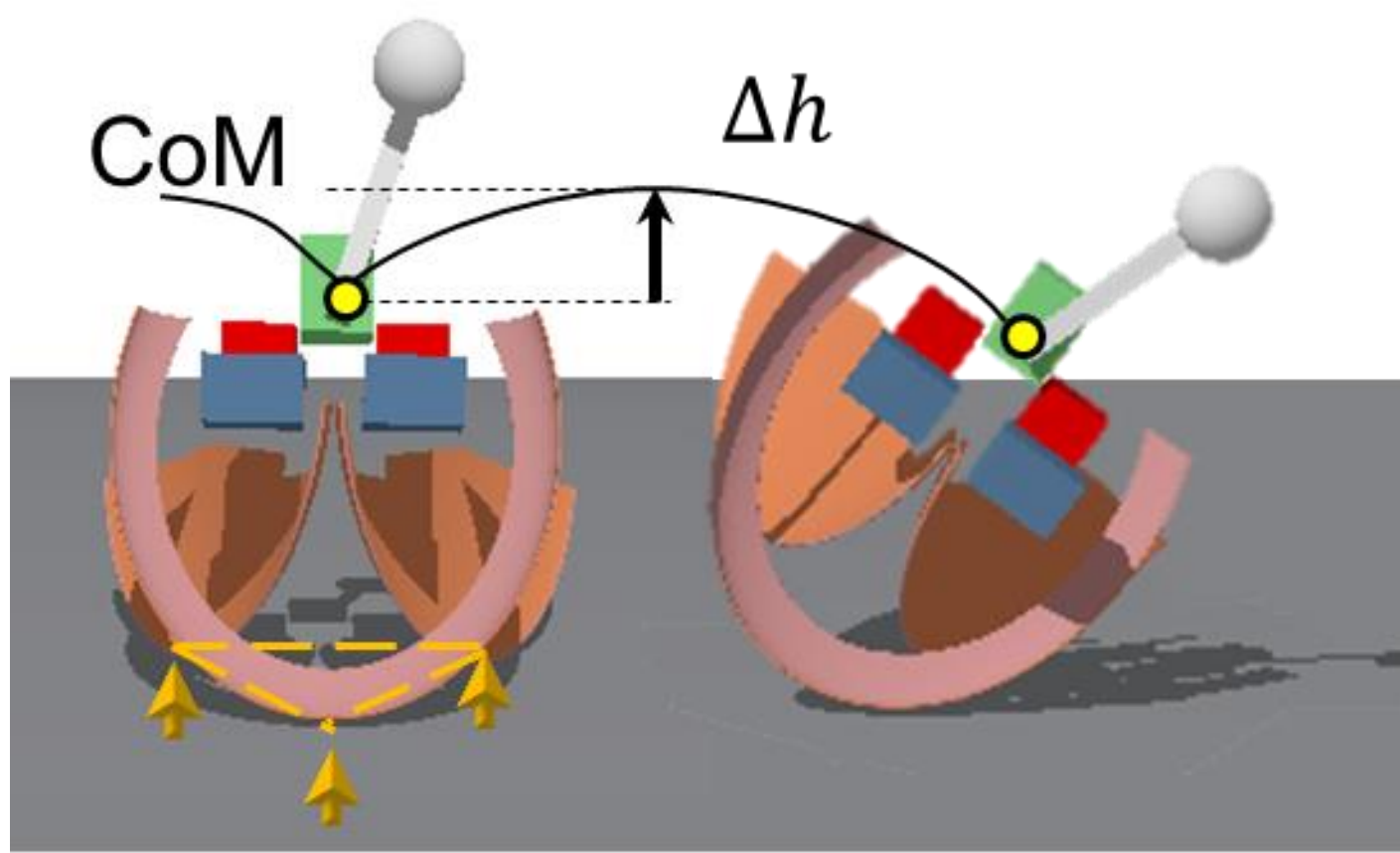

**Figure 3.** Snapshots of simulation robot during self-righting. Left: robot in metastable state, with two wings fully open and forming a triangular base of support (yellow arrows and triangle) with head on the ground. Right: robot pivoting over edge of head and right wing, overcoming the minimal potential energy barrier. Yellow dot is center of mass (CoM). Black curve is CoM trajectory. $\Delta h$ is the height lifted (exaggerated) and defines gravitational potential energy barrier $E_{\text{barrier}} = mg\Delta h$, where $m$ is robot mass and $g$ is gravitational acceleration.

## 3. Results

### 3.1. Comparison between simulation and physical robot experiments

Qualitatively, the simulation robot displayed similar self-righting motion as the physical robot (figure 4(a, b); supplementary video S2). In addition, the dependence of self-righting time on $\theta_w$ and $\theta_l$ was qualitatively similar between the two—it took both a shorter time to self-right as $\theta_w$ and/or $\theta_l$ increased (figure 4(c)). This qualitative similarity meant that the simulation had the fundamental physics correct.

However, for a given $\theta_w$ and $\theta_l$, it was easier for the physical robot to self-right than the simulation robot. This quantitative discrepancy was not surprising and likely due to several differences between the simulation and physical robots. First, the physical robot's thin wings and head likely deformed and decreased its potential energy barrier to self-right, compared to the simulation robot's rigid ones. In addition, the physical robot's left and right motors had small differences in actuation profiles (due to manufacturing variation). This lateral asymmetry may make it easier to self-right (Li et al., 2017). Furthermore, the Hertzian contact model used in simulation was developed for simple, ideal object shapes such as sphere and half-space (Popov, 2010). Contact mechanics modeling for objects with complex geometry is still an open area of research (Popov, 2010). Due to all these model approximations, quantitative match between the simulation and physical robots was difficult to achieve even after large scale parameter variation in simulation.

Our purpose is to study general principles of wing-leg coordination in leg-assisted, winged self-righting. Chrono Engine, as a matured physical engine, has been validated by some studies (Rieser et al., 2019; Tasora et al., 2016). Although there were quantitative discrepancies, our simulation still provided a useful tool to study the principles of how appendage coordination affected self-righting.

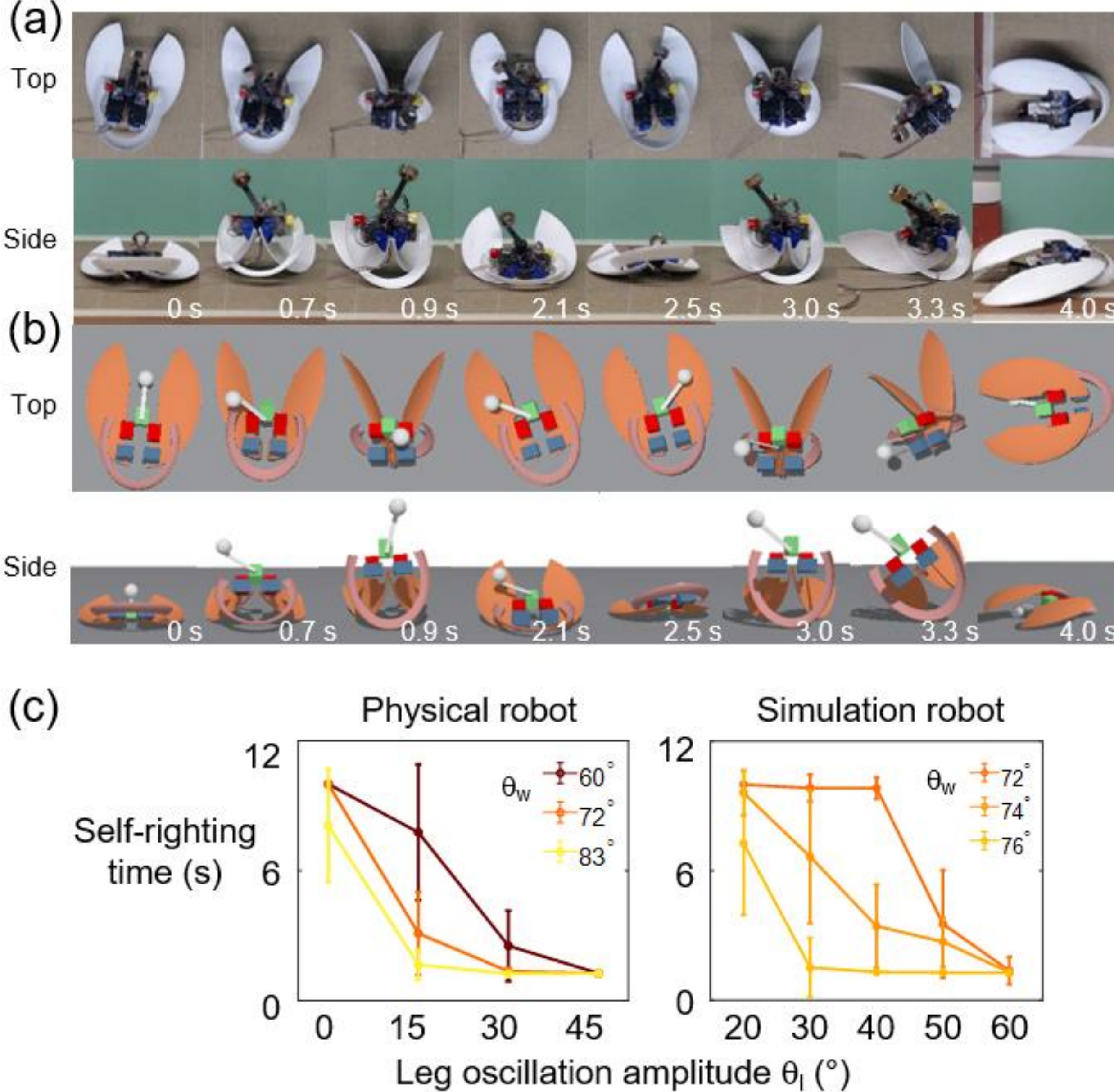

**Figure 4.** Validation of simulation robot against a physical robot. (a, b) Representative snapshots of physical (a) and simulation (b) robot experiments. (c) Self-righting time as a function of wing opening and leg oscillation amplitudes $\theta_w$ and $\theta_l$, comparing between physical and simulation robots. Error bar are ± s.d. $n$ = 5 trials at each combination of $\theta_w$ and $\theta_l$ for each robot.

### 3.2. Multiple attempts to self-right from metastable state

Similar to the animal (Li et al., 2019) and physical robot (Othayoth et al., 2017), the simulation robot's self-righting often required multiple wing opening attempts (figure 4(b); supplementary video S2).

If the simulation robot could not self-right upon the first attempt, it kept opening and closing its wings to make more attempts, until it either succeeded or failed to self-right within 10 s. This was because, like the animal and physical robot, the simulation robot was also often stuck in a metastable state with a triangular base of support, formed by the head and outer edges of two wings in contact with the ground (figure 2(a), yellow arrows and triangle). With sufficient perturbation from the leg, the simulation robot could escape from the metastable state, often after multiple attempts. These observations verified that our simulation robot displayed the strenuous leg-assisted, winged self-righting behavior that we are interested in.

### 3.3. Potential energy barrier

As wing opening amplitude $\theta_w$ increased, minimal gravitational potential energy barrier decreased (figure 5), because center of mass (CoM) height at the metastable state increased. As leg oscillation amplitude $\theta_l$ increased, minimal gravitational potential energy barrier also decreased (figure 5), because the CoM moved closer to the boundary of the triangular base of support (yellow triangle in figure 2(a)) when the leg rotated to one side. Thus, we should expect it to become easier for the robot to self-right as wing opening and leg oscillation amplitude increased, with everything else being equal.

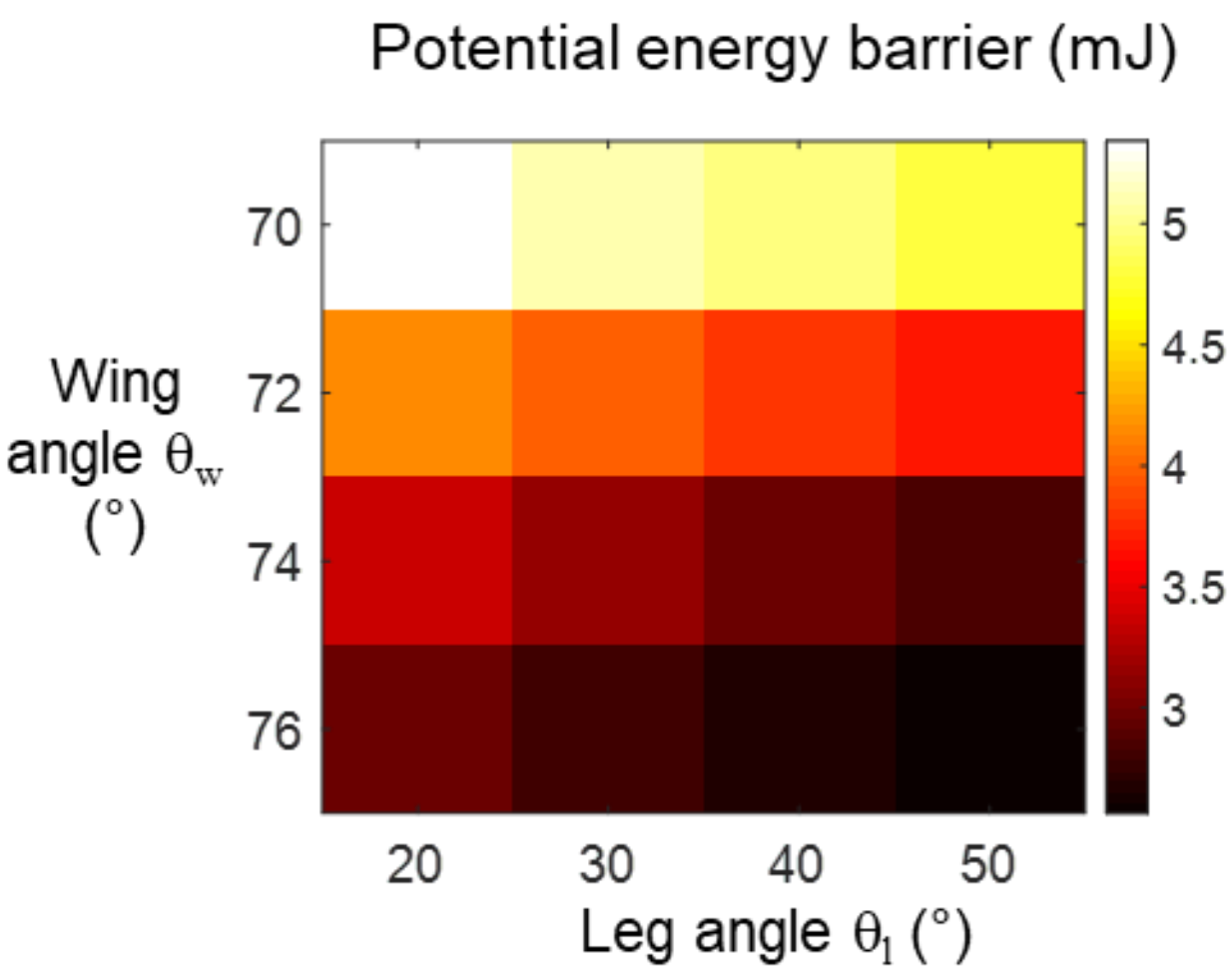

**Figure 5.** Minimal gravitational potential energy barrier as a function of wing and leg angles, which is the minimal energy barrier to escape the local potential energy minimum (Section 2.7).

### 3.4. Randomness in coordination increases self-righting probability

Without randomness in wing-leg coordination, self-righting outcome of the robot was deterministic. The robot either always succeeded or always failed to self-right for a given set of parameters (figure 6, $C_v = 0\%$). With randomness, self-righting outcome became stochastic (figure 6, $C_v > 0\%$). At each $C_v$, self-righting probability increased with wing opening and leg oscillation amplitudes (figure 6), as expected from the decreasing potential energy barrier (figure 5). Besides reducing the barrier, another reason that increasing leg oscillation amplitudes facilitated self-righting was that it increased leg rotation angular velocity and thus the kinetic energy that the leg injected.

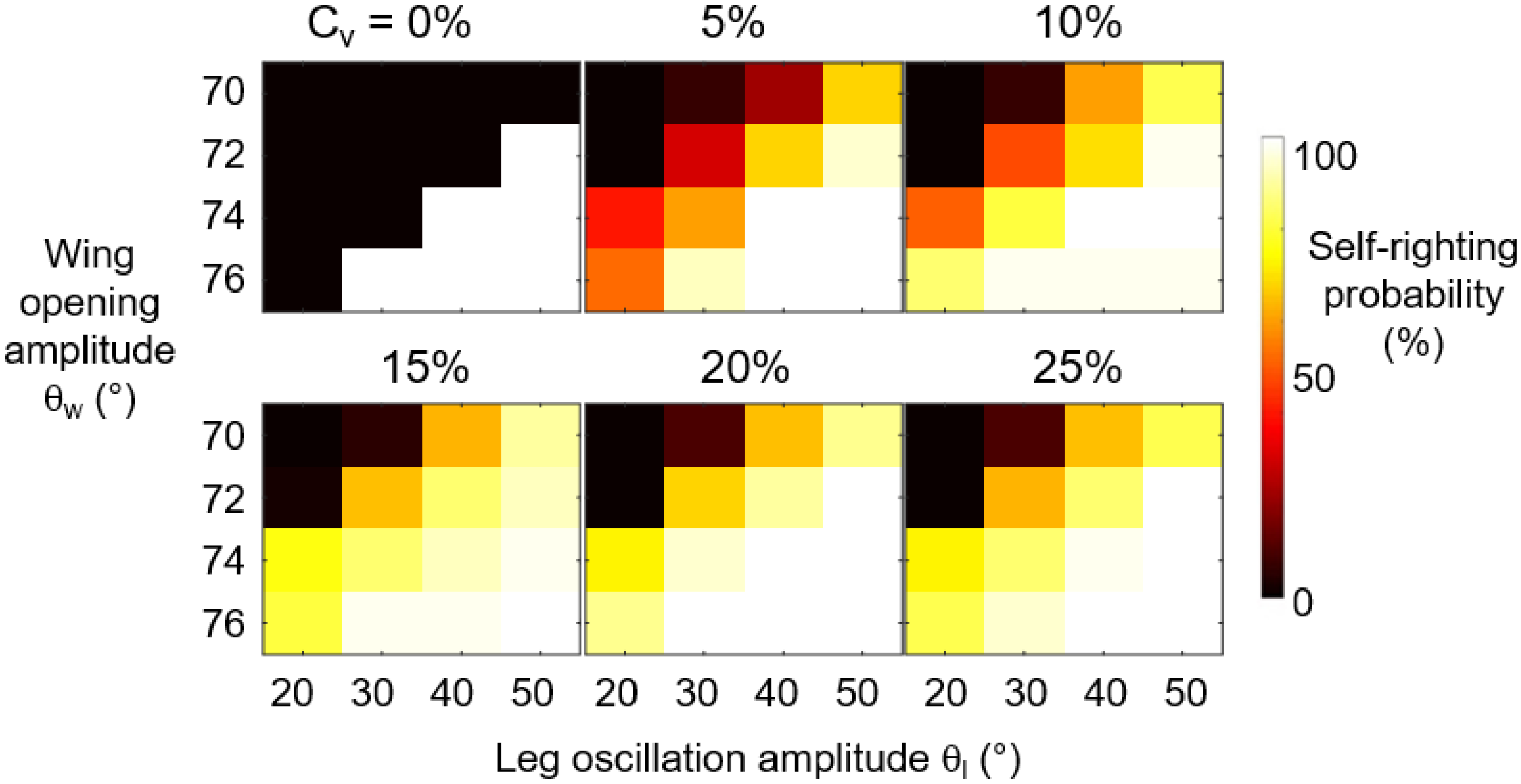

**Figure 6.** Effect of randomness in phase on self-righting probability within 10 s for each trial. Self-righting probability as a function of wing opening and leg oscillation amplitudes from simulation experiments, comparing across different levels of randomness $C_v$ in wing-leg coordination. Data shown are for initial phase $\varphi_1 = 70\%$. *n* = 40 trials at each combination of $\theta_w$ and $\theta_l$.

When $\theta_w$ and $\theta_l$ were too small or too large, self-righting probability was nearly always zero or one, not strongly affected by randomness (figure 6). However, at intermediate $\theta_w$ and $\theta_l$ near the boundary between success and failure without randomness ($C_v = 0\%$), increasing level of randomness in wing-leg coordination significantly increased the robot's self-righting probability (figure 6, $C_v = 0\text{-}25\%$). For $\theta_w$ and $\theta_l$ slightly above the boundary, probability decreased slightly, but this was outweighed by the substantial increase in probability for $\theta_w$ and $\theta_l$ slightly below the boundary.

These results suggested that, when a cockroach is too tired (weak wing pushing and/or leg flailing), randomness does not help; when it is very energetic (strong wing pushing and/or leg flailing), randomness does not matter. However, when an animal is nearly or barely able to self-right (intermediate wing pushing and leg flailing), which is frequent in strenuous self-righting, randomness in coordination significantly increases its chance of success. In addition, when randomness was sufficiently large, further increasing randomness did not significantly increase self-righting probability (figure 6, $C_v \geq 15\%$).

We note that the initial phase $\varphi_1$ shown in this example happened to be a bad phase ($\varphi_1 = 70\%$, figure 6; see definition in Section 3.5). However, if $\varphi_1$ happens to be a good phase, adding randomness may lead to requiring more attempts to self-right and thus decrease self-righting probability within a finite time.

### 3.5. Randomness in coordination changes phase between wing and leg oscillations

The phase between wing and leg oscillations had a strong impact on self-righting outcome at intermediate wing opening and leg oscillation amplitudes. This could be clearly seen from our results of self-righting outcome after a single attempt without randomness in coordination (figure 7, hereafter referred to as the single-attempt phase map). When $\theta_w$ and $\theta_l$ were sufficiently large (e.g., $\theta_w = 76°$, $\theta_l = 50°$), the simulation robot self-righted at nearly all $\varphi$. When $\theta_w$ and $\theta_l$ were sufficiently small (e.g., $\theta_w = 72°$, $\theta_l = 20°$), the robot never self-righted at any $\varphi$. When $\theta_w$ and $\theta_l$ were intermediate (e.g., $\theta_w = 76°$, $\theta_l = 20°$), phase became important. Empirically, some "good" phases led to success (e.g., phases around $\varphi = 0\%$, 50%, and 100%), whereas other "bad" phases led to failure (e.g., phases around $\varphi = 25\%$ and 80%).

The change of self-righting outcome from success to failure at higher wing opening or leg oscillation amplitudes (e.g., lower probability at $\theta_l = 50°$ than at $\theta_l = 40°$ for $\theta_w = 70°$, $\varphi = 40\%$) was likely because, besides changing the potential energy barrier, changes in wing opening and leg oscillation amplitudes also affected the energy injected and dissipated.

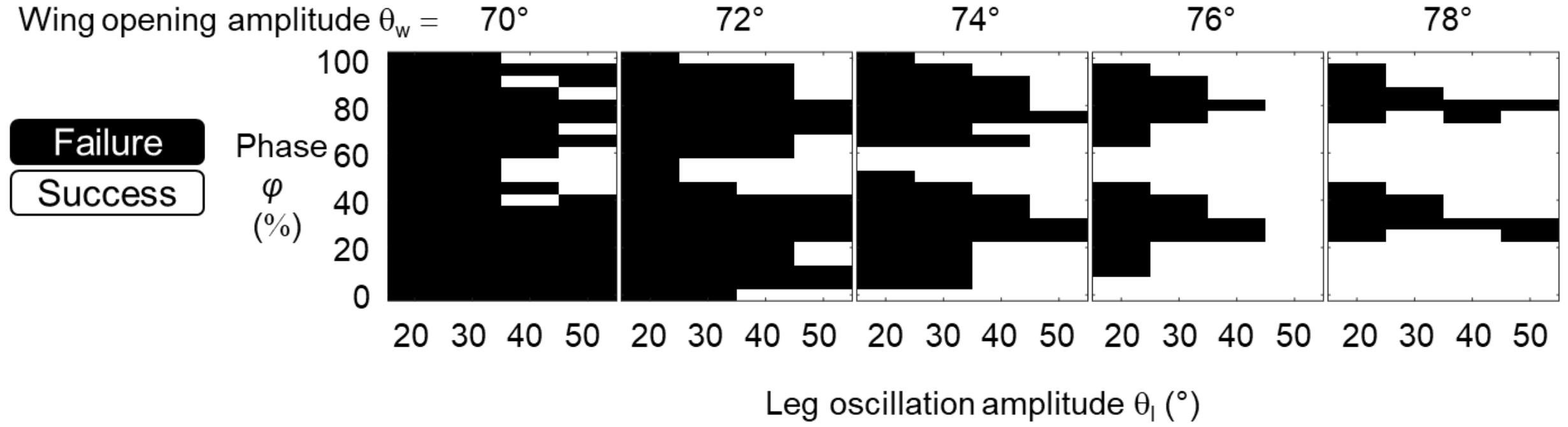

**Figure 7.** Dependence of self-righting outcome of a single attempt on phase (single-attempt phase map), without randomness in motion. Self-righting outcome as a function of φ and $\theta_l$, comparing across $\theta_w$.

### 3.6. Single attempt phase map predicts consecutive attempts outcome

After each failed attempt, the simulation robot oscillated little on the ground, because most of its kinetic energy was quickly dissipated (this was similar to the physical robot, see Section 3.1). Thus, consecutive attempts should be nearly independent of each other. Thus, we expected that the dependence of self-righting outcome on phase (phase map) was the same for every attempt. Then, we could use the phase map of a single attempt (figure 7) to predict the self-righting outcome after multiple consecutive attempts, by evaluating how phase evolved over attempts. If there is a strong history dependence between consecutive attempts, we cannot use the single-attempt phase map to make prediction, but we expect that randomness in coordination would still help because it extends the coordination space.

To test how well this worked, we predicted the number of attempts to achieve successful self-righting for various initial $\varphi_1$ and wing period $T_w$ without randomness (figure 8(a), right). Given initial $\varphi_1$, wing period $T_w$, and leg period $T_l$, we calculated how φ evolved for subsequent attempts (Section 3.7). We predicted that the attempt whose phase first reached the good phases in the single attempt phase map (Section 3.5) would be successful. Hereafter, we refer to this as the phase-based prediction method. The prediction of this method (figure 8(a), right) matched well with simulation results (figure 8(a), left) over the $T_w$ and $\varphi_1$ space. For a broader range of $T_w$ and $T_l$ (figure 8(b)), the phase-based prediction method achieved a high accuracy of 91 ± 6% (mean ± s.d.) in predicting the self-righting outcome observed in simulation experiments. We also used the phase-based method to predict self-righting probability with different levels of randomness $C_v$ (figure 6(c)), which well matched the results from simulation experiments (figure 8(c)).

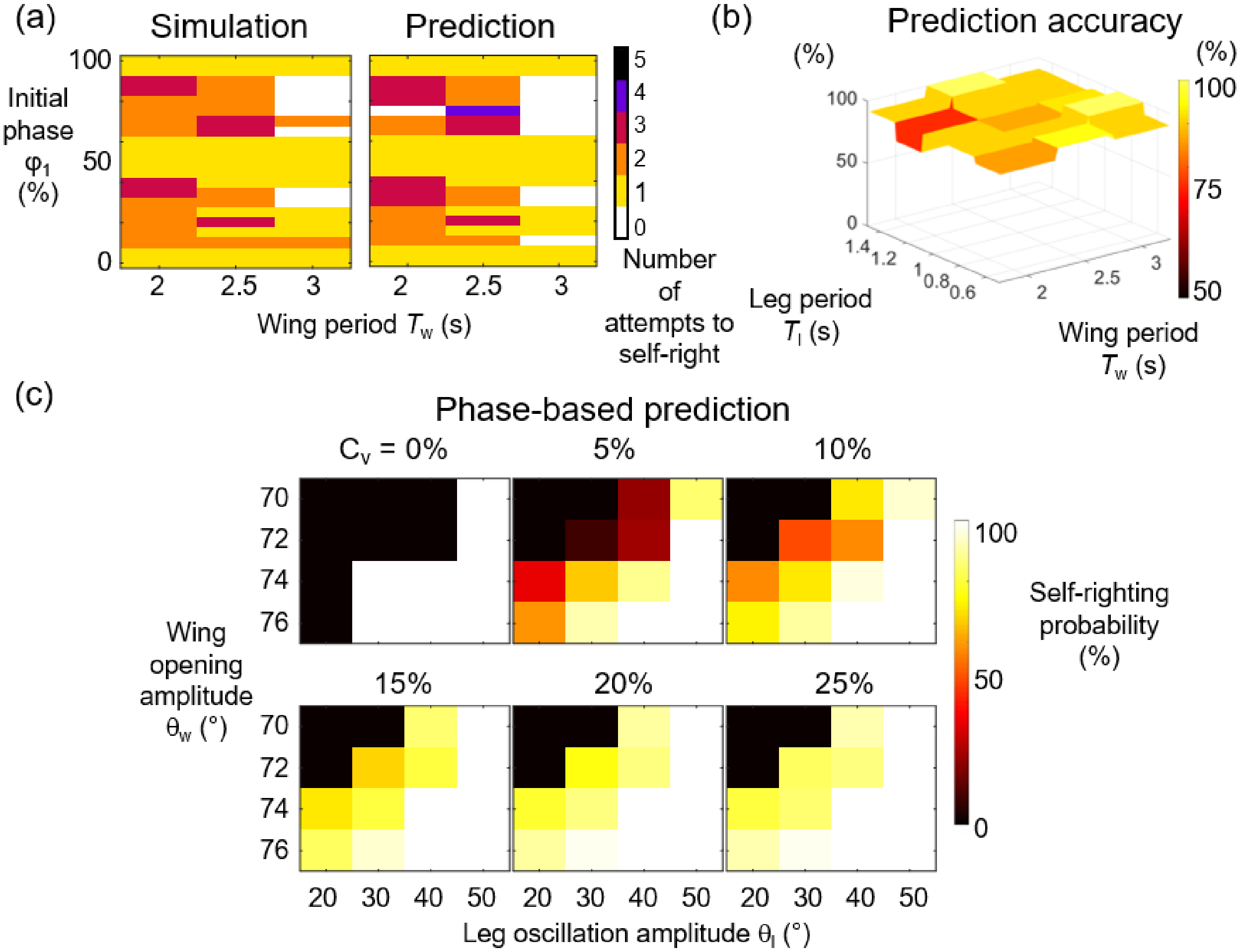

**Figure 8.** Using phase map from a single attempt to predict self-righting outcome after multiple attempts. (a) Prediction without randomness: Number of attempts required to self-right as a function of $\varphi_1$ and $T_w$, comparing between phase map prediction and simulation experiments. Failure to self-right is shown as number of attempts = 0. Data shown for $T_l = 0.6$ s, $\theta_w = 75°$, $\theta_l = 30°$. (b) Prediction without randomness: Prediction accuracy as a function of $T_l$ and $T_w$. Prediction accuracy is percentage of phases at which phase-based prediction matches simulation experiments. Data shown for $\theta_w = 75°$, $\theta_l = 30°$. (c) Prediction with randomness: Self-righting probability as a function of $\theta_w$ and $\theta_l$ predicted by phased-based method, comparing across randomness $C_v$ of coordination. $n = 40$ trials at each combination of $\theta_w$ and $\theta_l$.

### 3.7. Randomness allows stochastic visits of good phases

Because the phase of each attempt was a good predictor of its outcome, we examined how phase evolved over consecutive attempts to understand how randomness affected self-righting performance. Without randomness, wing and leg actuation were periodic. This resulted in a limited number of phases that could be visited during consecutive attempts. If the phases that could be visited happened to be bad for the wing and leg oscillation periods given, self-righting was never successful. For example, for wing period $T_w = 2$ s and leg period $T_l = 0.8$ s, the least common multiple of both periods was LCM(2, 0.8) = 4 s. Thus, the robot could only visit $n$ = LCM(2, 0.8)/2 = 2 different phases (e.g., figure 9(a, c) with an initial phase of $\varphi = 30\%$, followed by $\varphi = 80\%$, and so on and so forth). Both these two phases happened to be bad, and the robot was trapped in bad phases forever (supplementary video S3).

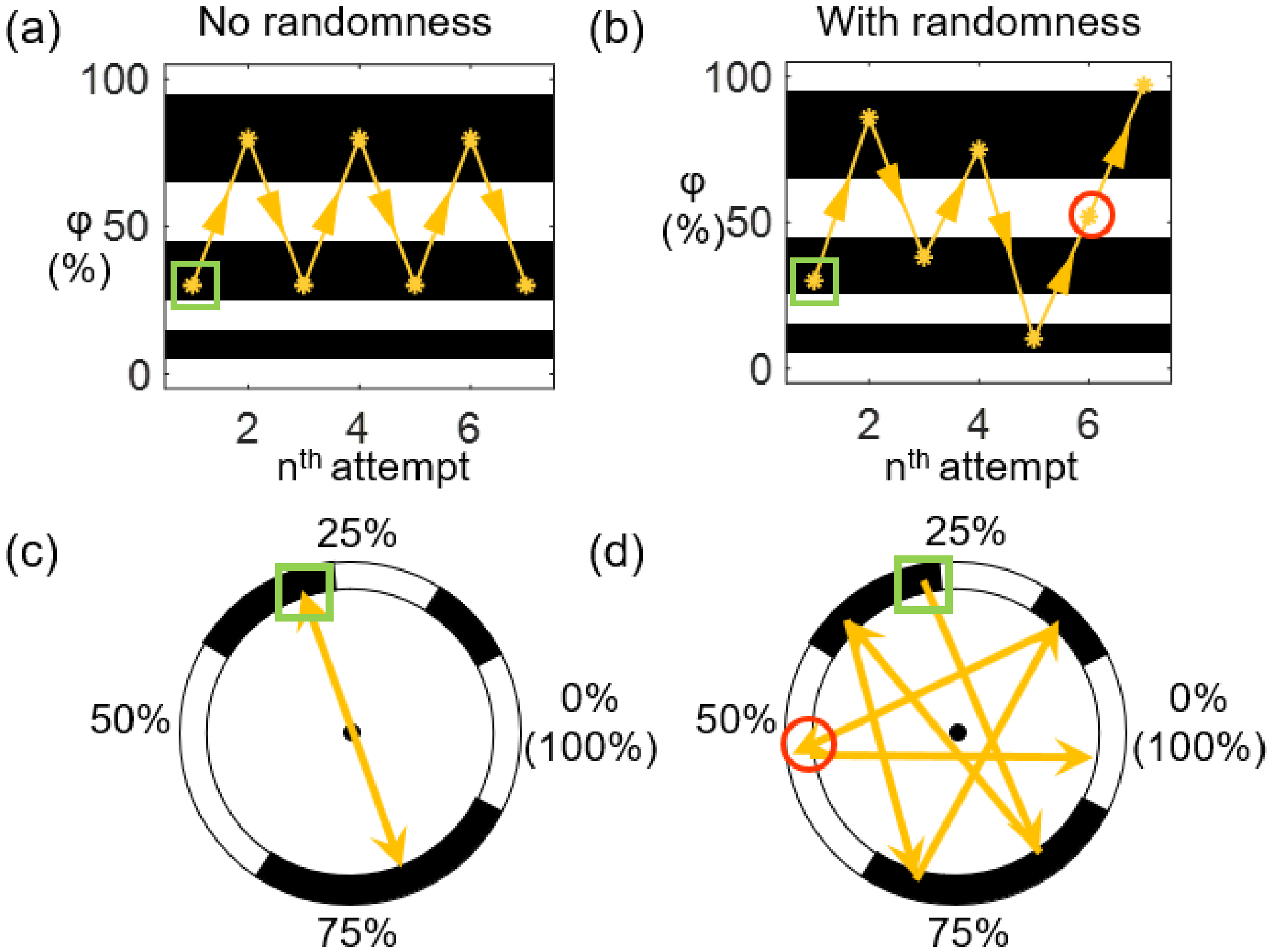

**Figure 9.** Phase evolution without and with randomness. Phase evolution over consecutive attempts (yellow points connected by arrows) on phase map (white: good phases, black: bad phases), without randomness (a, c) and with randomness (b, d). Phase map is from a single attempt in simulation experiments, assuming

that it does not change over attempts. Green box shows initial phase at the first attempt. Red circle shows the first good phase reached with randomness. Data shown for $\theta_w = 75°$, $\theta_l = 30°$, $T_w = 2$ s, $T_l = 0.8$ s.

With randomness added to the motion, the phase evolution was no longer periodic (figure 9(b, d)). Instead, the system could visit an infinite number of phases (any value from 0% to 100%) and was thus impossible to be trapped in bad phases, as long as there were good phases in the phase map (supplementary video S3). For the example case, as soon as it visited a good phase (the 6$^{th}$ attempt, figure 9(b, d)), the robot self-righted. This is true for all trials because the Gaussian noise that we added has a non-zero probability to reach any value.

## 4. Discussion

In summary, we studied the impact of randomness in coordination between appendages during strenuous, leg-assisted, winged self-righting. We developed a simulation robot following the design and control of a recent physical robot to generate the strenuous self-righting behavior, and we used it to conduct systematic simulation experiments and analyses. We discovered that randomness in wing-leg coordination facilitated self-righting, especially at intermediate wing opening and leg oscillation amplitudes, when the system's kinetic energy was about to overcome the potential energy barrier. Randomness allowed the system to explore various phases more thoroughly between wing and leg oscillations, thereby increasing the chance of finding a good coordination between them.

Although we did not systematically vary wing opening and leg oscillation periods, our phase-based prediction test results for different periods (figure 8a, b) indicated that randomness should facilitate self-righting by finding a good coordination even for different periods. Admittedly, the number of different phases accessible without randomness, $n = \mathrm{LCM}(T_w, T_l)/T_w$, can be made larger by choosing $T_w$ and $T_l$, which increases the chance of finding a good coordination. However, without randomness, the system still searches for a good phase over the phase space with a constant increment every attempt, which is a grid search over the phase space that covers only a finite number of phases. By contrast, adding randomness to the system results in a random search that is more thorough (can cover any phase) and efficient than a grid search (Bergstra and Bengio, 2012; Lerman, 1980).

Our work focused on the effect of randomness in its simplest form, in the phase between a pair of oscillating wings and a single leg, and only gave a glimpse into a very complex noisy system. Additional randomness may exist in the amplitudes, directions, and speed of the motions of multiple wings and legs, in the motions of other body parts such as the abdomen, as well as in the mechanical system itself (e.g., morphology, physical property, the environment), which may also be beneficial (see discussion in Section 4.2).

### 4.1. Implications for biological locomotion

Our results suggested that the large randomness in coordination during self-righting (as opposed to more periodic motion during walking and running) in the discoid cockroach and other species may be an adaptation to strenuous maneuvers. We speculate that animals may respond by moving their body and appendages more randomly when they encounter strenuous, emergency situations, such as being unable to self-right after multiple attempts, or becoming trapped by obstacles when moving in complex terrain (Gart and Li, 2018; Gart et al., 2018; Li et al., 2015; Othayoth et al., 2020).

Our study revealed the usefulness of randomness in biological and artificial system at the intermediate scale (body and appendage motion within a movement cycle), adding to previous knowledge at larger scales (e.g., trajectory over many body lengths and movement cycles) (Bénichou et al., 2005; Hoffmann, 1983; Reynolds and Rhodes, 2009) and smaller scales (e.g., sensory systems) (Gammaitoni and Bulsara, 2002; McDonnell and Ward, 2011; Wiesenfeld and Moss, 1995).

Our work complemented previous work on mitigating the negative impact of randomness to stabilize locomotion around limit cycles (Byl and Tedrake, 2009). In dynamic walking, randomness in terrain surface slope breaks the dynamic stability, which must be mitigated to maintain metastable locomotion (Byl and Tedrake, 2009). However, as we demonstrated, randomness in motion can also help escape being trapped in an undesired metastable state. This is especially useful if the locomotor task is strenuous. This insight may have broader implications. For example, when moving through complex terrain with large obstacles, animals and robots must often dynamically transition across distinct locomotor modes (Li et al., 2015; Othayoth et al., 2020). Our group's recent work demonstrated that, in different modes, their states are strongly attracted to different basins of an underlying potential energy landscape (Gart and Li, 2018; Han et al., 2017; Othayoth et al., 2020). Our study suggested that body and appendage coordination is crucial for quickly escaping from these attractive landscape basins and having large randomness in coordination is beneficial.

**4.2. Implications for robotics**

Our simulation robot differed from the physical robot in that it did not have noise in the mechanical system which is inevitable in the physical robot. Without random time delay in actuation, the result of simulation experiments is deterministic, whereas the result of the physical robot experiments is stochastic. Despite this difference, we expect that our conclusion also applies to physical robots, because randomness should help explore the coordination space and find a good coordination regardless. In fact, we expect having randomness in coordination to be even more useful for real, stochastic physical robots because, unlike the deterministic simulation for which good phases can be identified in advance, good coordination is unknown in a stochastic system and must be searched every time.

We speculate that randomness in coordination could also improve the performance of robots in other strenuous locomotor tasks. When robots are trapped in undesirable metastable states, such as in complex terrain (Gart and Li, 2018; Gart et al., 2018; Li et al., 2015; Othayoth et al., 2020), their normal gait (walking, running, etc.) may no longer work. Having or eliciting randomness in motions between body parts may help find a good coordination to escape from such unexpected emergencies.

Besides informing robot control, our discovery of the usefulness of randomness may also be useful for the mechanical design of self-righting robots. For example, one can use flexible and/or under-actuated appendages (e.g., using soft material and springs, or a hollow appendage with a heavy ball inside) to add stochasticity to the passive mechanics and dynamics.

**4.3. Future work**

Our study only discovered the usefulness of randomness but did not uncover the physical mechanism. Because successful dynamic self-righting requires sufficient mechanical energy to overcome potential energy barriers, wing and leg coordination influences self-righting probability presumably by changing the mechanical energy injected into system during each attempt. We are developing a simple template (analogous to (Libby et al., 2012; Patel and Braae, 2014; Saranli et al., 2004)) to model the hybrid dynamics of leg-assisted, winged self-righting to understand the physical mechanism (Xuan and Li, 2020). In addition, it would be interesting to test a suggestion from our study—that energetic animals should have less randomness in motion whereas fatigues animals should have more. Finally, it would be intriguing (but perhaps difficult (Heams, 2014)) to tease apart how much the randomness in animal self-righting (and other forms of strenuous locomotion) is uncontrolled and unintentional, and how much may be deliberate, controlled randomness as a form of behavioral adaptation.

**Acknowledgements**

We thank Ratan Othayoth, Shai Revzen, Noah Cowan, Simon Sponberg, Greg Chirikjian, Yang Ding, Avanti Athreya, and three anonymous reviewers for discussion and suggestions; Ratan Othayoth for providing data of physical robot experiments; Soowon Kim for early simulation development; Arman Pazouki and Dan Negrut for technical assistance on Chrono; and Zhiyi Ren for measuring Young's modulus.

This work was supported by an Army Research Office Young Investigator award (grant # W911NF-17-1-0346), a Burroughs Wellcome Fund Career Award at the Scientific Interface, and The Johns Hopkins University Whiting School of Engineering start-up funds to C.L.

## Appendix

### Measuring randomness level in animal wing/leg oscillation periods

As a measure of the level of randomness of the discoid cockroach's wing and leg motions, we chose to measure the coefficients of variation of the oscillation periods of the wings (both wings open and close simultaneously) and each hind leg. We analyzed the videos of three discoid cockroaches self-righting from (Li et al., 2019), with three trials for each individual. For each trial, we recorded the periods of all the oscillation cycles of both wings and each hind leg (figure 1(c)). We then separately pooled wing and leg oscillation period data from the three trials and calculated their respective mean and standard deviation for each individual (figure 1(d)) to obtain its coefficient of variations.

For wing oscillation, a cycle was defined as the interval between consecutive instants when the wings began to open and body began to pitch up. For leg oscillation, a cycle was defined to start when the tip of a hind leg reached the farthest position from the sagittal plane. We chose the hind legs because they are the longest and heaviest among the six legs (Kram et al., 1997) and contribute the most flailing kinetic energy (estimated to be 60% assuming similar angular velocities). Occasionally, the animal stopped moving its wings or legs for a while and then resumed moving. This pausing behavior had a disproportionately large effect on the variance of the periods. Thus, we excluded outliers of wing and leg oscillation periods that fall outside the interquartile range.

**Distribution of phase in animal motion**

Using animal data shown in figure 1D, we further calculated the animal's time delay (see definition in Section 2.1) and phase (see definition in Section 2.6) between wing and hind leg oscillation in each attempt. The range of phase in three individuals is from 0% to 100% (figure A1). This provided evidence that the animal's variations in wing opening and leg oscillation periods were sufficiently large for it to access a wide range of phases.

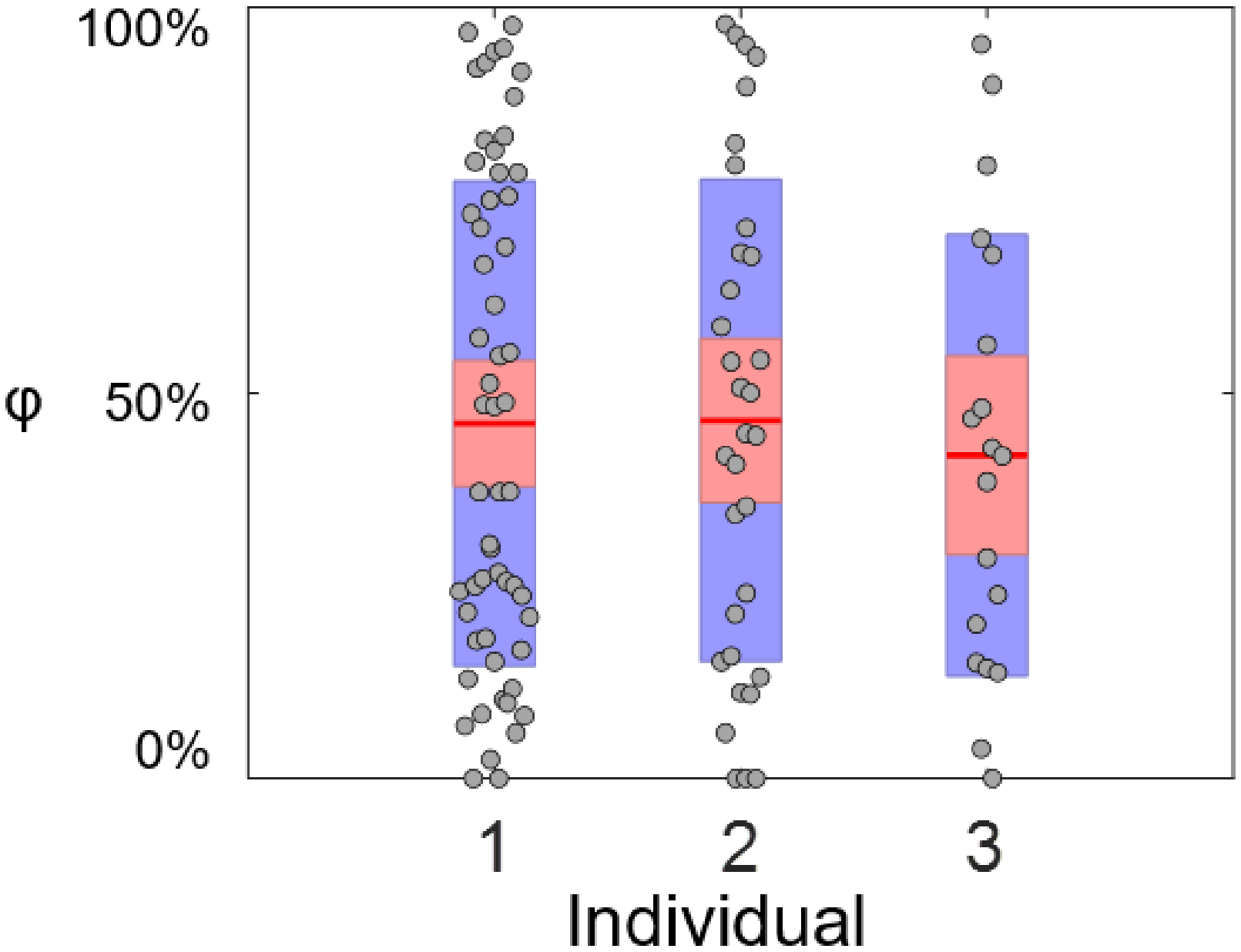

**Figure A1.** Distribution of phase in animal self-righting for three individuals ($n$ = 3 trials for each; 109 phases in total). Gray dots are phases for each attempt. Red line is mean. Pink rectangle shows 95% confidence interval of mean. Blue rectangle shows mean ± s.d.

## Supplementary videos

**Supplementary video S1.** Cockroach self-righting using wing pushing and leg flailing.

**Supplementary video S2.** Comparison of physical and simulation robot self-righting.

**Supplementary video S3.** Simulation robot self-righting with and without randomness in coordination. Top: simulation robot self-righting. Bottom: state trajectories in body pitch-roll space. Left: without randomness. Right: with randomness.